# Self-assembly dynamics of reconfigurable colloidal molecules

Indrani Chakraborty[†,§], Daniel J. G. Pearce[‡,⁑], Ruben W. Verweij[†], Sabine C. Matysik[†,§§],

Luca Giomi[‡], and Daniela J. Kraft*[,†]

[†] Soft Matter Physics, Huygens-Kamerlingh Onnes Laboratory, Leiden Institute of Physics, PO Box 9504, 2300 RA Leiden, The Netherlands.
[‡] Institute -Lorentz, Universiteit Leiden, PO Box 9506, 2300 RA Leiden, The Netherlands
* Corresponding author: kraft@physics.leidenuniv.nl

**Abstract:**

Colloidal molecules are designed to mimic their molecular analogues through their anisotropic shape and interactions. However, current experimental realizations are missing the structural flexibility present in real molecules thereby restricting their use as model systems. We overcome this limitation by assembling reconfigurable colloidal molecules from silica particles functionalized with mobile DNA linkers in high yields. We achieve this by steering the self-assembly pathway towards the formation of finite-sized clusters by employing high number ratios of particles functionalized with complementary DNA strands. The size ratio of the two species of particles provides control over the overall cluster size, or, "valence", as well as the degree of reconfigurability. The bond flexibility provided by the mobile linkers allows the successful assembly of colloidal clusters with the geometrically expected maximum valence. We quantitatively examine the self-assembly dynamics of these flexible colloidal molecules by a combination of experiments, molecular dynamics simulations, and an analytical model. Our "flexible colloidal molecules" are exciting building blocks for investigating and exploiting the self-assembly of complex hierarchical structures, photonic crystals and colloidal meta-materials.

Colloidal particles are powerful model systems to study the assembly and phase behaviour of atoms and molecules.[1] Already the simplest model, a colloidal sphere with isotropic interaction potential, has provided unprecedented insights into complex and fundamental processes such as crystallization,[2,3] the glass transition,[4,5] and melting/fracture.[6,7] To capture the anisotropic interactions and shapes of more complex molecules, colloidal constructs made up of multiple "colloidal atoms" are being developed and further functionalized with site-specific interactions by various approaches.[8] In analogy with their molecular cousins, they are called "colloidal molecules"[9–11] and are expected to revolutionize the field of material science due to their tunable complexity and additional control over the assembly process. However, with a few exceptions, their syntheses consist of many step processes.[12–16] The simpler techniques often produce a wide range of colloidal molecules that require time-consuming separation protocols before they can be employed in experiments,[17,18] and the low yield for certain valences limits their usability in experiments.[18,19] Furthermore, despite their shape and interaction analogy, all current realizations of colloidal molecules consist of rigid objects that fail to mimic the structural flexibility present in real molecules. Yet, structural reconfigurability is essentially the property that has been predicted to strongly affect the assembly and phase behaviour of colloidal molecules, and any rigid system will have limited use as a model system.[20]

Recent theoretical studies on colloidal molecules with such structural reconfigurability have found a variety of unusual behaviors, such as the appearance of novel crystalline lattices[21,22] and other ordered structures, that are kinetically inaccessible for rigid colloidal molecules.[23] For instance, hierarchically assembled flexible colloidal molecules, have been numerically shown to give rise to thermodynamically stable liquid-liquid phase separation.[22] This long sought-after phase transition may explain the origin of the anomalies of liquid water, but is difficult to probe experimentally. With increasing bond flexibility, the crystalline phase has been furthermore predicted to be replaced by a previously unexpected state of matter, the fully bonded disordered network state.[23] To date, none of these predictions have been experimentally validated because of the rigidity of current experimental realizations of colloidal molecules.

Here we exploit colloidal elements that provide flexibility between linked particles, so called colloidal joints,[24,25] to assemble different types of flexible colloidal molecules with high fidelity and yield. We achieve this by choosing a high number ratio of two sets of spheres functionalized with complementary, surface-mobile DNA strands and assemble them into small clusters, where one sphere type surrounds spheres of the other type.[26–28] Depending on their size ratio and thus packing density of the spheres on the outside, colloidal molecules with different degrees of flexibility and "valence", i.e. cluster size, can be realized. We investigate the growth dynamics of colloidal molecules with different valences using experiments and Molecular Dynamics (MD) simulations and describe it quantitatively by a theoretical model that considers the availability of bonding space on the surface of the central particle. The high yield combined with a tunable flexibility and valence of our flexible colloidal molecules not only makes them an unprecedented model system for studying the phase behaviour of molecules, but also exciting building blocks for creating reconfigurable materials or bits in wet computing. [28–30]

RESULTS AND DISCUSSIONS

**Assembly strategy**

To create flexible colloidal molecules we follow a strategy based on assembling spherical particles onto the surface of a central sphere.[26,31] The size ratio $\alpha = R_i/R_o$ of the two spheres determines the cluster geometry by simple packing arguments, where $R_i$ and $R_o$ are the radius of the inner and the outer sphere, respectively.[26,32] To steer the assembly pathway towards finite size clusters, the outer particle species is used in excess of the particle species intended to form the core of the cluster, see Figure 1A-B. Until now, the experimental realization of this straightforward idea, however, has struggled with what has been termed the "random parking" problem: when two bonding particles cannot rearrange, which is the case for interactions based on surface-bound DNA linkers[31] and charge,[31,33,34] the optimal packing of the outer particles cannot be achieved.[31] This precluded the formation of the geometrically predicted clusters and led to clusters with non-uniform shapes and sizes. Even the flexible bonds between lock-and-key shaped particles mediated by depletion interactions could not achieve the assembly of only one type of cluster due to kinetic barriers.[35] The assembly of a single flexible tetrahedral cluster by holographic optical tweezers was demonstrated as a proof of principle.[28] However, while it is very useful to study the dynamics of individual colloidal molecules with internal degrees of freedom,[20,28,36] the time-consuming particle-by-particle addition is not a viable approach for the production of more than a few individual flexible colloidal molecules.

To resolve the "random parking" problem and overcome kinetic barriers, we use colloidal joints as the central particle of the cluster.[24,25] Colloidal joints are particles that provide the same structural flexibility between linked colloids as their macroscopic analogues by exploiting DNA linkers that are mobile on the particle surface.[37] Experimentally, we realize colloidal joints by coating silica particles of diameters 1.15 ± 0.05 μm, 2.06 ± 0.05 μm, 3.0 ± 0.25 μm, and 7.0 ± 0.3 μm with a closed lipid bilayer into which DNA linker strands with double cholesterol moieties are anchored (Figure 1A). At room temperature, the bilayer is in the liquid phase (the transition temperature $T_m$ of DOPC is −17° C),[38] allowing free diffusion of the cholesterol-anchored DNA linkers on the particle surface (Figure 1A). See Methods for experimental details. Colloidal particles connected to DNA linkers that are mobile on the surface inherit their ability to freely diffuse over the surface of the colloidal joint particle (Figure 1B). This surface-mobility enables us to obtain clusters with maximum packing (Figure 1C) as the outer particles can *rearrange* after binding and thereby provide access to the core particle surface for additional oncoming particles until saturation and hence the geometrically expected valence has been reached. At the same time, it is the crucial element to create internal degrees of freedom in the colloidal molecules. By choosing DNA linkers with 11 base pair long single stranded sticky ends we ensure essentially irreversible binding even at high packing densities and overcome any kinetic barriers.

By combining flexible joints functionalized with DNA linker strands *S'* (represented by green particles in Figure 1A-C) with an excess of particles functionalized with the complementary strand *S* (represented by magenta particles in Figure 1A-C) we obtain finite-sized colloidal clusters with full flexibility of the attached outer spheres, or, reconfigurable colloidal molecules (see Methods section). We can experimentally realize quasi-2D (Figure 1D) or 3D (Figure 1E) flexible colloidal molecules by simple selection of the material of the outer spheres and its associated gravitational height in water. When we use solid silica spheres with surface-mobile DNA linkers $S$ as the outer particles of the cluster, the experiment is confined to quasi-2D, see Figure 1D and supporting movie S1. Employing lower-density polystyrene beads with DNA linkers $S$ relieves this constraint and leads to three-dimensional flexible colloidal molecules as shown in Figure 1E and supporting movie S2. Because a high-quality lipid bilayer relies on silica surfaces, we used polystyrene beads with surface-bound DNA linkers. For simplicity, we restrict ourselves to the quasi-2D case in the following experiments and discussions.

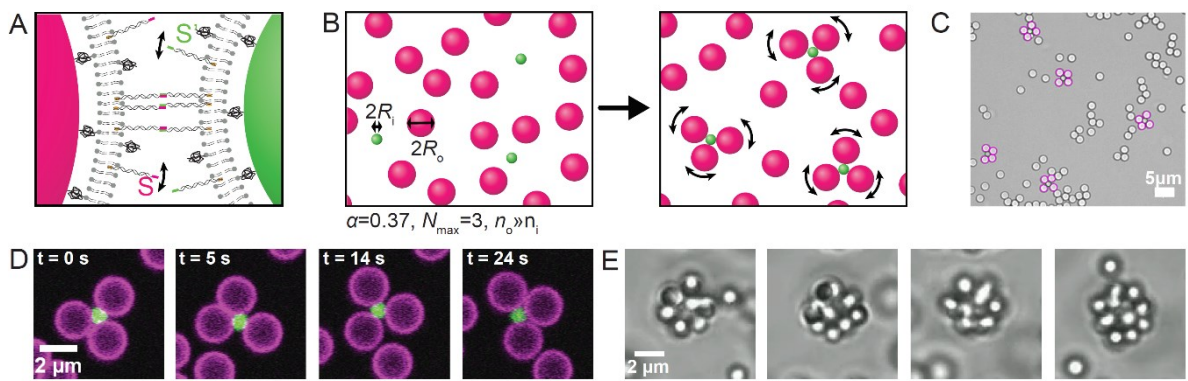

**Figure 1. Self-assembly of reconfigurable colloidal molecules.** A) Schematic diagram of the self-assembly process. Combining two types of colloidal spheres ($S$ and $S'$) functionalized with complementary DNA linkers (indicated in magenta and green) at high number ratios leads to the formation of finite size clusters. The flexible bond is formed through linkages between surface mobile DNA linkers anchored into a lipid bilayer on the particle surface. B) Schematic showing colloidal spheres with radius ratio $\alpha = 0.37$ and maximum valence $N_{max} = 3$, assembling into colloidal molecules when the density of outer spheres, $n_o$, is much greater than that of the inner spheres, $n_i$. The surface mobility of the DNA linkers imparts reconfigurability to the bonded particles. C) Bright-field microscopy image of the self-assembled colloidal molecules for $\alpha = 0.67$. Clusters with $N_{max} = 4$ are highlighted by colored outlines. Scalebar is 5 μm. D) Time-lapse confocal images of a quasi-2D reconfigurable colloidal molecule composed of 2.06 ± 0.05 μm silica particles (magenta) surrounding a central 1.15 ± 0.05 μm silica particle (green). E) Time-lapse bright field images of a 3D colloidal molecule composed of 1.15 ± 0.05 μm polystyrene particles encompassing a central 2.06 ± 0.05 μm silica particle. The lower density of the polystyrene particles enables the polystyrene particles to diffuse on the curved surface of the central sphere.

**Reconfigurable colloidal molecules with valence tunable by geometry**

The size and thus valence of our reconfigurable colloidal molecules can be straightforwardly controlled through the choice of the size ratio $\alpha = R_i/R_o$. A colloidal molecule with $N$ outer spheres is considered to have valence $N$. For a closed packed combination of hard spheres in 2D, the maximum valence $N_{max}$ of the corresponding 2D cluster is given by[26,39]

$$N_{max} = \frac{\pi}{\arcsin\left(\frac{1}{1+\alpha}\right)} \tag{1}$$

purely from geometrical considerations. In a physical system, $N_{max}$ can only have integer values, whereas the fractional part of $N_{max}$ accounts for gaps in between the spheres. Taking the integer part of $N_{max}$ yields a stair curve[26] with increasing $\alpha$, see Figure 2A. We demonstrate the assembly of flexible colloidal molecules with valences $N_{max} = 3$ to $N_{max} = 7$ by making various combinations of differently sized spheres as outer or inner particles of the colloidal molecules. The results are displayed in Figure 2B, with symbols denoting the respective size ratios in Figure 2A. For example, for a radius ratio of $\alpha = 0.29$ we use an excess of $7.0 \pm 0.3$ μm spheres in combination with few $2.06 \pm 0.05$ μm spheres to obtain clusters with maximum valence $N_{max} = 3$, see Figure 2A-B. In contrast to previous attempts at assembling clusters through this approach, we here find that the experimentally obtained maximum valences $N_{max}$ perfectly agrees with the values expected on the basis of the size ratio of the constituent spheres. This is the direct consequence of the reconfigurability of the system, which solves the random parking problem and allows the assembly of many different types of flexible colloidal molecules with the expected maximum valences.

Interestingly, each cluster size $N_{max}$ can be formed within a range of size ratios $\alpha$. This is represented by the finite width of each step for every integer value of $N_{max}$ (Figure 2A). At the lower end of the size ratio for any given $N_{max}$, the outer spheres are closely packed, while at the higher end there is sufficient space to almost fit an additional particle and thus more area is available to move. Therefore, the choice of the size ratio $\alpha$ also provides a handle to control the range of angular motion of the colloidal molecules and hence their flexibility. We note that the term "flexibility" has been used to describe both the angular motion range or the angular speed of the outer particles. Angular motion range and angular speed may both contribute to the intuitive meaning of flexibility and we hence use it in this way here. To show that the size ratio of the two sphere types enables a tunable motion range, we assemble colloidal molecules with maximum valence $N_{max} = 4$ from spheres with size ratio $\alpha = 0.5$ and $\alpha = 0.67$. Visual inspection of the resulting colloidal molecules confirms that the outer spheres of $\alpha = 0.67$ colloidal molecules have more space to move on the surface of the central particle than they do for $\alpha = 0.5$ colloidal molecules, see Supporting Movie S3. We quantify the angular motion range of the two types of $N_{max} = 4$ colloidal molecules by measuring the probability distribution $P(\theta)$ of the angle $\theta$ between any two outer spheres and the central particle (Figure 2C), when the maximum valence has been attained. Colloidal molecules with $\alpha = 0.67$ show a wider spread in the angular distribution indicating a higher degree of flexibility, while the angular motion of the spheres in the more closely packed $\alpha = 0.5$ cluster is more constrained. In addition, the minimum angle that is geometrically possible as well as the most probable angle shifts towards larger angles with decreasing size ratio due to the larger outer particles, pointing yet again towards a stronger confinement of the outer spheres for smaller size ratios.

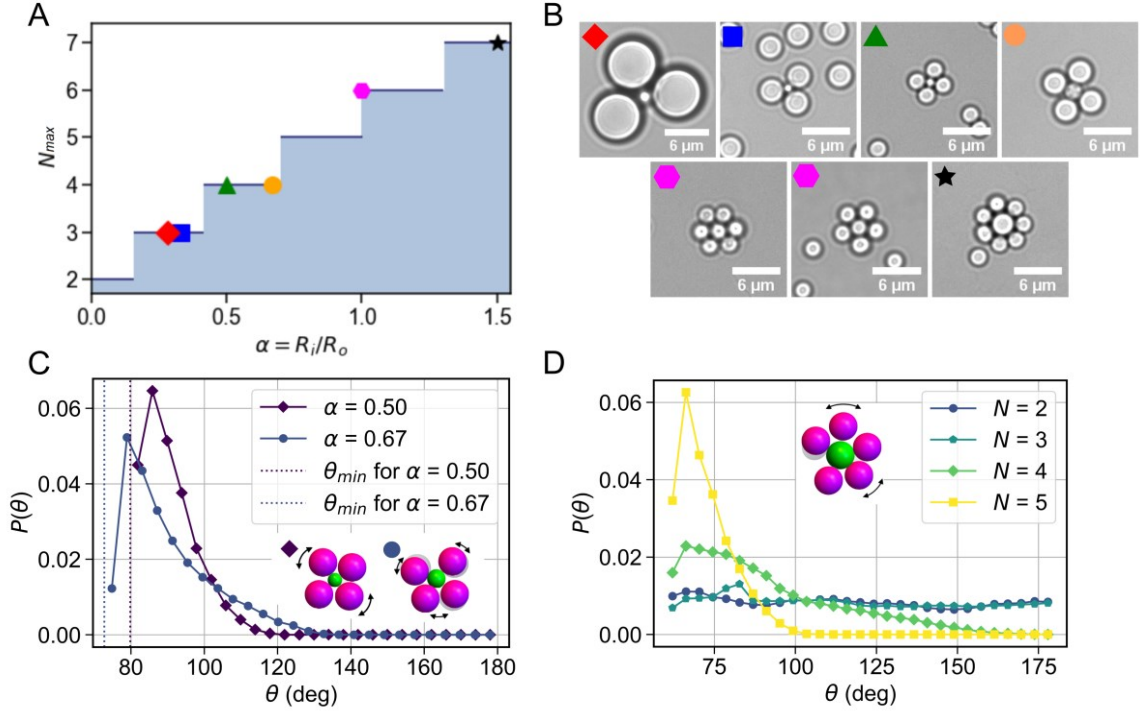

**Figure 2. Reconfigurable colloidal molecules of different size ratios and corresponding valences.** A) Geometrically expected maximum valence $N_{max}$ for different size ratios $\alpha$ and B) corresponding bright field images of obtained colloidal molecules at select ratios: ◆ $\alpha = 0.28$, $R_i = 2.06 \pm 0.05$ µm, $R_o = 7.0 \pm 0.3$ µm; ■ $\alpha = 0.33$, $R_i = 1.15 \pm 0.05$ µm, $R_o = 3.0 \pm 0.25$ µm; ▲ $\alpha = 0.5$, $R_i = 1.15 \pm 0.05$ µm, $R_o = 2.06 \pm 0.05$ µm; ● $\alpha = 0.67$, $R_i = 2.06 \pm 0.05$ µm, $R_o = 3.0 \pm 0.25$ µm; ⬢ $\alpha = 1$, $R_i = 2.06 \pm 0.05$ µm, $R_o = 2.06 \pm 0.05$ µm; ★ $\alpha = 1.5$, $R_i = 3.0 \pm 0.25$ µm, $R_o = 2.06 \pm 0.05$ µm. Here, $\alpha = R_i/R_o$ where $R_i$ and $R_o$ are the radii of the inner (core) and outer particles, respectively. For $\alpha = 1$ the majority of colloidal molecules had valence $N = 5$, and occasionally $N = 6$ was observed. C) Probability distribution $P(\theta)$ of the angle $\theta$ between any two adjacent outer spheres in a colloidal molecule for $\alpha = 0.5$ and $\alpha = 0.67$, shows that the angular motion range of the outer spheres for a given maximum valence ($N_{max} = 4$ here) is tunable through the size ratio and is larger for higher values of $\alpha$. Inset shows schematics of the two resulting clusters (not to scale to demonstrate different available space). D) The angular motion range decreases as the valence increases as can be seen from $P(\theta)$ for $\alpha = 1$ for valences $N = 2, 3, 4, 5$.

The motion range also decreases with increasing valence during the assembly process. To investigate this quantitatively, we plot $P(\theta)$ for $\alpha = 1$ ($N_{max} = 6$) for increasing valence from $N = 2$ to $N = 5$ (Figure 2D). While flexible colloidal molecules with $N = 2$ and $N = 3$ show a flat angular distribution, a peak appears for $N = 4$ that becomes pronounced at $N = 5$. This shift in the angular distribution is again connected to the increasingly constrained motion of the outer spheres. We note that this confinement may also affect the speed with which the outer spheres can move over the surface of the central particle.[20] Because the distances between the outer particles have an effect on their angular speed through hydrodynamic interactions, the flexibility is expected to decrease as a function of increasing valence due to both a smaller angular range as well as lower angular speed. Control over the flexibility is particularly exciting in view of the predicted impact on the phase diagram of limited valence particles.[23] Tuning of the size ratio of the constituent particles thus enables us to obtain a wide variety of colloidal molecules with both controlled valence and tunable flexibility.

**Dynamics of colloidal molecule growth**

To reach high yields of colloidal molecules with maximum valences, we need to get a better understanding of the dynamics of their self-assembly process. We examine this process by a combination of experiments and MD simulations. For the experiments, we chose: a) $1.15 \pm 0.05$ µm silica spheres as the core particle and $2.06 \pm 0.05$ µm silica spheres as the outer particles and b) $2.06 \pm$

0.05 μm silica spheres as the core particle and 3.0 ± 0.25 μm silica spheres as the outer particles, both cases corresponding to an expected maximum valence of $N_{max}$ = 4. The particles were mixed in a number ratio of 1:5 and 1:20 to obtain good statistics and the growth of the colloidal molecules was observed for 1.2 days. At this number ratio, a few colloidal polymers[24,36,40] and other composite structures were produced besides the colloidal molecules, which were excluded from the total count to enable comparison with the simulations. The population growth at different valences including and up to the expected maximum valence $N_{max}$ = 4 were recorded at fixed time intervals.

We complemented these experiments with molecular dynamics simulations to examine the self-assembly process for a range of values of *α*. The colloidal spheres were approximated as two-dimensional soft disks with radii $R_i$ and $R_o$ which represent the central and outer particles, respectively; hence $R_i=\alpha R_o$ (Figure 3A). Each disk undergoes an uncorrelated random walk with constant step length *T*. When two simulated disks come into contact, the distance between the centers of the two disks is less than the sum of their radii, $R_{ij} < R_i+R_o$. This results in a small overlap *d* which we use to define a repulsive force $F_R = K_{disk}\,d$ directed away from each other (Figure 3B), where $K_{disk}$ is a constant. Each disk is assigned a certain type of DNA tether, *S* (or *S'*), with a rest length $R_t$ and a maximum length $R_{tmax}$. If the distance separating two disks with complementary DNA tethers is less than $R_i+R_o+2R_t$, then a bond is formed between these disks (Figure 3B). This bond behaves like a linear spring when $R_{ij} > R_i+R_o+2R_t$, i.e. there is an attractive force $F_A = max\ [K_{\text{tether}} \times (R_{ij} - R_i - R_o - 2R_t),\ 0]$, with $K_{\text{tether}}$ a constant representing the spring stiffness. This force is merely an attractive force having no torsional effect on the disk, and replicates the mobility of the DNA linkers used in experiments. If the distance between two bound disks exceeds $R_i+R_o+2R_{tmax}$ the bond between them is broken, implying the bonds have a finite strength. The resulting energy landscape for bonded disks contains a wide minimum representing the range of separation between two bonded disks at which they apply no forces to each other. This is essentially some additional 'wiggle room' provided by the non-zero length of the DNA linker. All simulations were performed with a single particle of radius $R_i$ surrounded by a bath of 99 particles of radius $R_o$ (Figure 3C). Simulations were initialised by placing particles at random, non-overlapping positions in a periodic box of length *L*. We scale all lengths by

$R_o$ and control $R_i$ by adjusting $\alpha$ and fix $R_t = 0.02$. We set $L = \sqrt{9(99 + \alpha^2)}$ giving a constant packing fraction of $\pi/9$. The remaining parameters are set to $K_{disk} = K_{tether} = 1$, $R_{tmax} = 10R_t$ and $T=0.01$. Reported values were obtained by averaging over 500 simulations all with the number ratio 1:99.

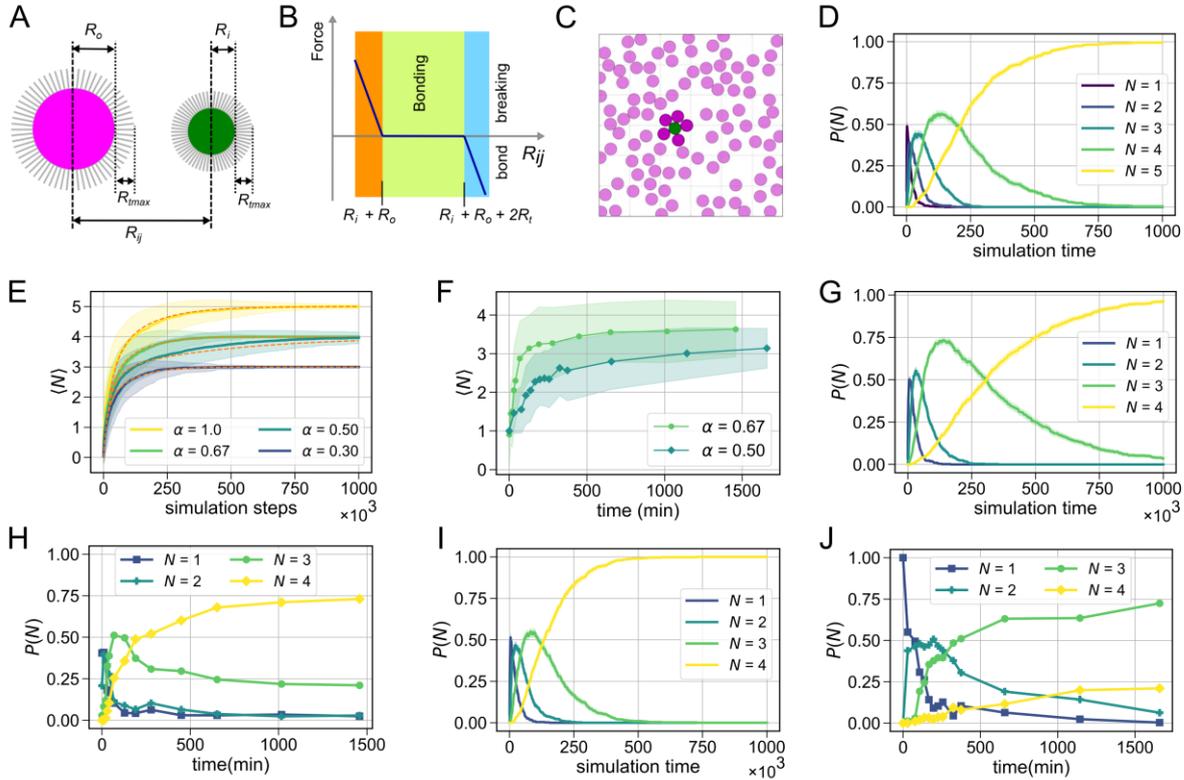

**Figure 3. Comparison between experiments and Molecular Dynamics (MD) simulations of the self-assembly dynamics of a colloidal molecule.** A) Schematic diagram of the simulation setup. Two disks with radii $R_i$ and $R_o$ each have linkers of maximum length $R_{tmax}$ on their surface, and interact with B) a distance dependent force. The force is repulsive when $R_{ij} < R_i + R_o$, it is zero in the range $R_i + R_o < R_{ij} < R_i + R_o + 2R_t$ and attractive when $R_i + R_o + 2R_t < R_{ij} < R_i + R_o + 2R_{tmax}$ until bond breaking occurs for $R_{ij} > R_i + R_o + 2R_{tmax}$. C) Representative still from the simulation of a reconfigurable colloidal molecule for $\alpha = 1.0$. The core particle (green) is surrounded by an excess of complementary particles (magenta). D) Simulated probability $P(N)$ of finding a cluster with valence $N$ as a function of time for $\alpha = 1$ shows that assembly saturates at $N = N_{max} - 1$. E) Simulated (solid line) and calculated (orange dashed line) ensemble averaged valence $\langle N \rangle$ for $\alpha = 0.3$, $\alpha = 0.5$, $\alpha = 0.67$ and $\alpha = 1.0$ as a function of simulation steps show excellent agreement. F) Experimentally obtained ensemble average valence $\langle N \rangle$ for $\alpha = 0.5$ and $\alpha = 0.67$, respectively, as a function of time. G) Simulated and H) experimentally measured $P(N)$ for $\alpha = 0.67$. I) Simulated and J) experimentally measured $P(N)$ as a function of time for $\alpha = 0.5$.

We expect that the available space on the surface of the central particle not only influences the flexibility but also directly influences the speed with which the clusters reach full saturation. Intuitively, the assembly should proceed faster if more space is available on the central particle for binding another outer particle. This is clearly observed in the simulated probability *P(N)* for all colloidal molecules that have bound *N* particles for *α* = 1.0 (Figure 3D). As we go from *N* = 1 to *N* = 5, the peak becomes broader and broader, indicating a slowdown of the growth process. We quantitatively investigate this by comparing the evolution of the expected valence of the central cluster ⟨*N*⟩ for four values of *α*, namely *α* = 0.30, *α* = 0.50, *α* = 0.67, and *α* = 1.0 in simulations (Figure 3E). The average valence of the clusters increases monotonically with time towards the maximum allowed number. Initially growth is fast and then slows down while saturating at a maximum valence. The same behaviour of ⟨*N*⟩ is also obtained in our experiments for *α* = 0.5 and *α* = 0.67, see Figure 3F. An increasing size ratio for the same maximum valence $N_{max}$ indeed leads to a speed up of the assembly process, see *α* = 0.5 and *α* = 0.67 for $N_{max}$ = 4. The flexibility and assembly speed and, ultimately, the yield, are thus directly related to the size ratio of the two spheres.

The maximum valence shown in Figure 3E agrees with the geometrically predicted one for *α* = 0.30, *α* = 0.50, and *α* = 0.67, but, surprisingly, only reaches *N* = 5 instead of the expected *N*$_{max}$ = 6 for *α* = 1.0. We observed a similar behaviour in the experiments, where the majority of colloidal molecules for *α* = 1.0 features maximum valence 5 instead of 6. For this reason, we show examples of both *N* = 5 and *N*$_{max}$ = 6 clusters for *α* = 1.0 in Figure 2B. A careful look at the bright field image in Figure 2B provides insight as to why colloidal molecules in experiments occasionally reach *N*$_{max}$ = 6: the outer spheres have slightly different scattering patterns indicating a variation in height from the surface. While simulations are restricted to 2D, the quasi-2D nature of the experiments allows out-of-plane diffusion. This together with a (small) size polydispersity of the colloidal particle creates additional space to access and bind to the core particles, albeit with low probability. The majority of clusters, however, saturates at a valence *N* = *N*$_{max}$−1 for *α* = 1.0. This behaviour can be understood from an entropic argument: to achieve *N*$_{max}$, the outer particles need to be tightly packed around the core particle in a single state, leaving no space for internal motion. This state is entropically unfavourable compared to the large number of non-close-packed arrangements that are possible due to the internal flexibility, see also the distribution of angles between outer particles shown in Figure 2C. This sub-maximum saturation behaviour occurs for all valences at the smallest size ratio at which they still can be assembled, or, in other words, just after the transition from one maximum valence to the next higher one.

To further compare experiments and simulations we looked at the assembly dynamics for $N_{max} = 4$ and $α = 0.5$ and $α = 0.67$. For this, we plot the probability $P(N)$ for all colloidal molecules that have bound $N$ particles for both simulations (Figure 3G and I) and experiments (3H and J). The binding of the first particle occurs very fast, and all core particles quickly have one particle bound. This still leaves ample surface available for binding a second particle, and indeed, leading to a sharp peak of $P(1)$ and a quick increase of $P(2)$. The addition of the third particle shows a slight slow-down, but still occurs at a rate comparable with that characterizing the binding of the first two particles, consistent with the almost $θ$-independent angular motion range shown in Figure 2D. With the significantly reduced available space on the surface and the surface mobility of the bound particles which hinder access to this surface further, the binding of the final particle now clearly slows down. In the experiments for $α = 0.67$, 71% of the core particles attain a minimum valence $N = 4$ within 1000 min (Figure 3H), whereas only 17.4% attain $N = 4$ for $α = 0.5$ within 1000 min (Figure 3J). Note that the assembly is faster for the $α = 0.67$ case despite the slow diffusion of the larger outer particles (3 μm diameter) as compared to the case of $α = 0.5$ (outer particles having 2 μm diameter). This clearly indicates that once the core particles are surrounded by an excess of outer particles, the dominating factor that controls the speed of the assembly is the available space on the core particle surface, which in turn is decided by the size ratio $α$.

Despite the gradual slowdown of the assembly process, a reasonably good yield of flexible colloidal molecules with maximum valences could be obtained simply by introducing reconfigurability to the system. Non-reconfigurable clusters typically yield distributions of cluster sizes due to the random distribution of the outer spheres, which can only be circumvented for $N_{max} = 2$ and $N_{max} = 4$ by choosing a specific $α$.[31] Here, in this work, the reconfigurability in principle enables optimization for any size ratio and maximum valence. However, the self-assembly of colloidal molecules with a given $N_{max}$ progresses faster for larger size ratios: for $\alpha = 0.5$, we obtained 21% of $N_{max} = 4$ colloidal molecules in a time interval of 1.2 days while for $\alpha = 0.67$ it was 73% in a time interval of 1.0 day, see Figure 3J and H, respectively. The yields could be further improved by using inert DNA or PEG linkers[25] on the colloidal joints to passivate any remnant non-specific interactions in the system and by a thorough passivation of the bottom glass surface of the sample chamber so as to let the self-assembly process continue unhindered over days. Altogether, our simple model captures the behaviour of the experimental system well.

**Analytic model for the assembly dynamics of flexible colloidal molecules**

We now attempt to turn our hypothesis that the self-assembly dynamics are governed by the available surface area on the core particle into an analytical model. For this, we assume that the formation of the colloidal molecules is a diffusion-limited process of successive collisions between the central *S'* particle (green) with incident *S* particles (magenta). Incident particles have probability $p_N(\alpha)$ of forming a bond with a cluster already containing *N*−1 particles and with size ratio *α*. Using an independent Poisson process to model the addition of particles, the probability of having added the *N*th particle to a cluster currently containing *N*−1 particles in a time interval *δt* can be described by the cumulative density function (CDF),

$$C(N, \delta t) = 1 - e^{-\frac{p_N(\alpha)\delta t}{t_0}}. \qquad (2)$$

Figure 4A shows the cumulative probability (simulated) of adding the *N*th particle to a cluster for *α* = *1.0* (solid lines), plotted alongside equation (2) (orange dashed line), in which we have used regression to fit for the value of $p_N(\alpha)/t_0$. By combining the CDFs (2) for subsequent cluster sizes with the corresponding probability density functions (PDFs), i.e. their time derivatives, we obtain an analytic expression for the expected valence as a function of time. Using the values of $p_N(\alpha)/t_0$ found by regression, this can be used to predict the average growth of a cluster for a range of values of *α,* Figure 4B (see SI and Figure S1 for details).

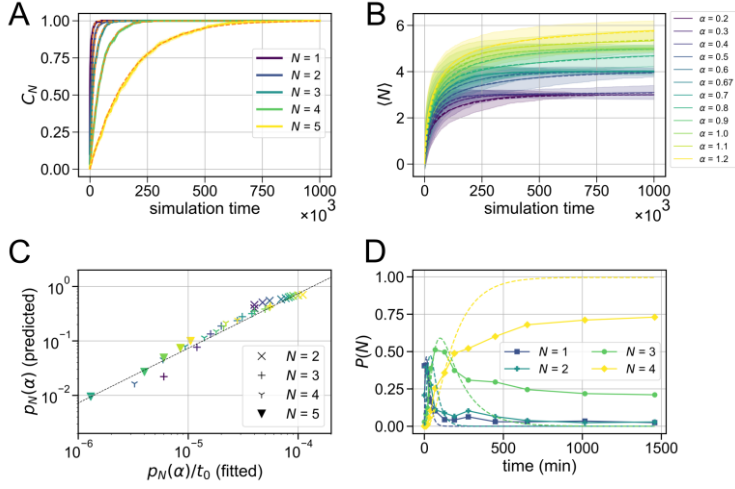

**Figure 4. Analytical model of the assembly of a reconfigurable colloidal molecule.** A) Cumulative probability (simulated) of adding the $N^{th}$ particle to a cluster for $\alpha = 1.0$ (solid lines), plotted alongside equation (2) (orange dashed lines), in which we have used regression to fit for the value of $p_N(\alpha)/t_0$. B) Average growth (simulated) of the valence of a cluster for a set of values of $\alpha$ ranging from $\alpha = 0.2$ to $\alpha = 1.2$ (solid lines) along with analytically obtained curves obtained using the values of $p_N(\alpha)/t_0$ found by regression (dashed lines), see SI. C) Comparison of $p_N(\alpha)/t_0$ obtained by regression to $p_N(\alpha)$ predicted by the entropic theory. The dashed line shows a linear relationship with $t_0 = 7.5 \times 10^3$. D) Fit of the analytical model (dashed line) to the experimentally obtained $P(N)$ vs time plots (solid lines) for $\alpha = 0.67$. While the fit is good at shorter time intervals, it diverges at longer times.

Since the system is isotropic, we can assume that an incident particle can arrive from any direction, and therefore also estimate the values of $p_N(\alpha)$ as the fraction of configurations that allow for the addition of an $N^{th}$ particle to a cluster with $N-1$ bound particles. To calculate this, we describe a cluster by the angles between the bound particles. Without loss of generality, we fix the angular position of the first bound particle to be $\theta_1 = -\varphi$, where $\varphi$ is the minimum angle allowed between two particles shown in supporting Figure S2 and given by $\cos(\varphi) = 1 - 2R_0^2(R_i + R_o + 2R_t)^{-2}$. The angular positions of the remaining particles are then labeled sequentially in a clockwise manner. The available angular space for a second particle is then simply $2\pi - 2\varphi$ and thus $p_2(\alpha) = 1 - \varphi/\pi$. The addition of a third particle will now depend on the position of the second particle. If $\theta_2 \approx 0$ there will be available space for the addition of a third particle in the region $\theta_3 \in [\theta_2 + \varphi, 2\pi - 2\varphi]$, however if $\theta_2 \approx 2\pi - 2\varphi$ the third particle will only have space within the region $\theta_3 \in [0, \theta_2 - \varphi]$. To account for all possible configurations, we must integrate both scenarios over all possible positions of $\theta_2$ to find $p_3(\alpha) = [3\varphi - 2\pi]^2/[2\pi(2\pi - 2\varphi)]$ (see SI and Figure S3). A similar process can be followed for every subsequent particle that is added, detailed in the SI. When comparing the estimates of $p_N(\alpha)/t_0$ obtained via regression to those predicted by this entropic theory we find a linear relationship, see Figure 4C. By performing a linear fit in Figure 4C, we can estimate the simulation value of $t_0 = 7.5 \times 10^3$. Furthermore, we can show that the characteristic time $t_0$ scales linearly with the particle radius, $R_o$, inversely with the squared packing fraction, $\Phi^2$, and inversely with the squared step length, $T^2$, in the simulation, $1/t_0 \sim T^2\Phi^2/R_0$ (see SI). We plot the prediction of the aforementioned model (orange dashed lines) alongside the simulation results (solid lines) with the single fitting parameter $t_0 = 7.5 \times 10^3$ in Figure 3E.

We can now match our analytic model to the experimental results for $\alpha = 0.67$. To do so, we experimentally measured the probabilities $P(N)$ for a single flexible colloidal molecule to have *exact* valence $N$ as a function of time. We subsequently made least square fits to this data using the above expressions for $p_2$, $p_3$ and $p_4$. As shown in Figure 4D, we found good overall agreement for $\alpha = 0.67$ at shorter time intervals. The deviation at longer time intervals might stem from a slowdown in the assembly process due to non-specific interactions between the particle surface and the bottom glass surface of the sample chamber or due to degeneration of the lipid bilayer over several hours. It has been recently shown that by inserting lipopolymers of required molecular weights or double stranded inert DNA linkers in the lipid bilayers, this problem of non-specific interactions over extended time periods can be solved to a large extent.[25]

The assembly dynamics of flexible colloidal molecules can be summarized as being fast initially and then gradually slowing down with increasing valence. This behaviour can be explained by the fact that since there are no long-range attractive forces present in the system, an oncoming particle has to come in close proximity to the central particle and stay there for a sufficiently long time so as to form a linkage. Therefore, a smaller availability of space on the core particle leads to a lower probability of attaining the geometrically expected maximum valence in a given time interval.

CONCLUSIONS

Reconfigurable colloidal molecules were assembled by combining two types of spherical colloids functionalized with complementary surface mobile DNA linkers in high number ratios. The size ratio of the constituent spheres determines the formation of flexible colloidal molecules with different geometrically determined valences. The obtained valences for each size ratio matched with the predicted maximum valences for a closed packed configuration of hard spheres in 2D. The reconfigurability of the system thus solves the random parking problem that long hindered the formation of densely packed structures in earlier experiments. Additionally, the size ratio also determines the motion range of the colloidal molecules which allowed us to tune the flexibility of the colloidal molecules. High yields of the geometrically predicted maximum valence can be achieved for almost any size ratio and valence, although entropic effects make attaining close packed configurations impossible. The growth dynamics of the colloidal molecules were examined in detail using experiments and molecular dynamics simulations, as well as analytic calculations. The assembly rate was observed to become slower as the colloidal molecules approached maximum valence, owing to increasingly less available space on the core particles. For the same reason, smaller size ratios for a given valence led to slower self-assembly. The experimental data showed good qualitative agreements with molecular dynamics simulations and the analytical model. The high yields for any valence and internal degrees of freedom make reconfigurable colloidal molecules an exciting model system for studying the behaviour of flexible (bio)molecules such as intrinsically disordered proteins, immunoglobulins or enzymes.[41–43] They have great potential to be utilized as the basic units for assembling complex, hierarchical structures, photonic crystals and colloidal metamaterials and as model systems to study phase transitions by tuning the valences and the motion ranges.

METHODS

**Materials.** Silica colloids (1.15 ± 0.05 μm, 2.06 ± 0.05 μm, 3.0 ± 0.25 μm, 7.0 ± 0.29 μm) were obtained commercially from Microparticles GmbH or synthesized in the laboratory by a modified Stöber's method.[44] The lipids 1,2 dioleoyl-*sn*-glycero-3-phosphocholine (DOPC), 1,2-dioleoyl-*sn*-glycero-3-phosphoethanolamine-N-[methoxy(polyethylene glycol)-2000](ammonium salt) (DOPE-PEG$_{2000}$), 1,2-dioleoyl-*sn*-glycero-3-phosphoethanolamine-N-(lissamine rhodamine B sulfonyl)

(ammonium salt) (DOPE-Rhodamine) and 1,2-dioleoyl-sn-glycero-3-phosphoethanolamine-N-(carboxyfluorescein) (ammonium salt) (DOPE-Fluorescein) were obtained at >99% purity from Avanti Polar Lipids, Inc. Three different DNA strands (Eurogentec) with the following sequences were used: a) A strand: Cholesterol-TEG-5'-TTT-ATC-GCT-ACC-CTT-CGC-ACA-GTC-AAT-CTA-GAG-AGC-CCT-GCC-TTA-CGA-*GTA-GAA-GTA-GG*-3'-6FAM b) B strand: Cholesterol-TEG-5'-TTT-ATC-GCT-ACC-CTT-CGC-ACA-GTC-AAT-CTA-GAG-AGC-CCT-GCC-TTA-CGA-*CCT-ACT-TCT-AC*-3'-Cy3 and c) C strand: Cholesterol-TEG-3'-TTT-TAG-CGA-TGG-GAA-GCG-TGT-CAG-TTA-GAT-CTC-TCG-GGA-CGG-AAT-GC-5'. The A and B strands consist of a backbone that can be hybridized with the C-strand, and an 11 base pair long sticky end with complementary sequences at the 3' ends (denoted by italic characters). To identify the different linkers, A strands are labeled with the fluorescent dye 6-FAM, i.e. 6-Carboxyfluorescein (excitation: 488 nm, emission: 521 nm, depicted in green) and B is labeled with the fluorescent dye Cy3 (excitation: 561 nm, emission: 570 nm, depicted in magenta), respectively.

**Preparation of silica particles with surface mobile DNA linkers.** We functionalized silica particles with surface mobile DNA linkers, using a protocol suggested by van der Meulen et al.[37] At first, small unilamellar lipid vesicles (SUVs) were prepared from a mixture of DOPC and DOPE-PEG$_{2000}$ in a 90:10 molar ratio in chloroform. If necessary, for increasing fluorescence, 0.001% mole fraction of DOPE-Rhodamine or DOPE-Fluorescein was added to label the membranes. The lipids were desiccated in vacuum, and re-suspended in HEPES buffer (10 mM HEPES, 47 mM NaCl, 3 mM NaN$_3$, pH = 7.01) for 30 mins to obtain a 3g/L solution. This was followed by 21 times extrusion of the lipid mixture through two stacked polycarbonate filters (Whatman) with 30 nm pore size to obtain SUVs. 50 µM solutions of the A or B DNA strand were hybridized with the C strand in a 1:1.5 volume ratio by heating the solution to 90°C and cooling down at 1°C/min. The hybridized DNA strands consisted of a 47 base pair long double-stranded central part, with double cholesterol anchors connected through TEG (tetraethylene glycol) spacers at one end and an 11 base pair long single-stranded sticky part at the other end which could link to a complementary sticky end. *S'* strand was produced by hybridizing the A and C strand while *S* was produced by hybridizing the B and C strand.

At first the SUVs were added to an equal volume of 5g/L solution of silica particles and put on a rotating tumbler for 40 mins at a slow turn speed of 9 rotations/min, in order to prevent sedimentation due to gravity. Then the particles were centrifuged for 5 min at 494 rcf and washed with HEPES. Required amounts of DNA were added to the SUV encapsulated particles and the resulting solution was again kept on the turner for 1 hour. Finally, to get rid of any excess dye or lipids, the particles were washed three times in HEPES by centrifugation.

**Formation of Colloidal Molecules.** To produce the colloidal molecules, two groups of particles with similar or different sizes and functionalized with complementary DNA, were mixed together in a number ratio such that the designated core particles were surrounded by an excess of the other particles. We note that number ratios below 1:3 tend to produce a mixture of different geometries including colloidal molecules, colloidal polymers and clusters without any definite configurations. At number ratios above 1:8 mostly and above 1:20 only colloidal molecules were obtained.

**Sample Observation.** Flexible colloidal molecules were imaged in sample holders with a hydrophobized and passivated glass as the bottom surface. The glass surface was hydrophobized using Surfasil (a siliconizing agent), and passivated by adding 5% w/v Pluronic F-127 to the holder, storing it for 30 mins and rinsing with water. An inverted Nikon TI-E microscope equipped with an 100x objective lens (NA = 1.4) was used for both confocal and bright field imaging. Images were recorded using A1R confocal scanhead in 8kHz resonant scan mode equipped with GaAsp detectors in confocal and a monochrome CCD camera (DS-QiMc) in bright field mode. Excitation was achieved using a 40mW Argon laser (488nm) and a 20mW sapphire laser (561nm).

**Molecular dynamics simulations**. Simulations following the algorithm outlined in the text were solved iteratively over time. To approach maximum valence 100 disks were placed in a periodic boundary at random locations (with no overlaps). 99 of these was selected to have radius $R_0$ and the remaining ones have radius $\alpha R_0$; we set $R_0 = 1$ and use this to define our lengthscale. The tethers were given length $R_t = 0.02 R_0$, which is consistent with typical lengthscales in experiments. The system size is rescaled to give a constant packing fraction of $\pi/9$ for all values of $\alpha$. The size of steps in the random walk were $T = 0.01$ for all particles. The spring constants associated with collisions and stretching of the DNA tether, $K_{disc}$ and $K_{tether}$ respectively, were both set to 1, representing very stiff springs. The system was simulated for $10^6$ timesteps, long simulations were necessary due to the low probabilitiy of reaching high valence. All simulation results presented here are the average of 500 independent realisations of the model.

AUTHOR INFORMATION

IC and RV planned and executed experiments, DJGP designed and executed simulations, DJGP and LG developed the analytical model, IC, RWV, DJGP and DJK analyzed data, DJK conceived the project, all authors contributed to the writing of the manuscript.

**Corresponding Author**

*Kraft@Physics.LeidenUniv.nl

**Present Address:**

§Department of Physics, Birla Institute of Technology and Science, Pilani - K K Birla Goa Campus, Zuarinagar, 403726, Goa, India

‡Dept. of Mathematics, Massachusetts Institute of Technology, 182 Memorial Dr, Cambridge, Massachusetts, USA
Dept. of Theoretical Physics, University of Geneva, Quai Ernest Ansermet 30, Geneva, Switzerland

§§Yusuf Hamied Department of Chemistry, University of Cambridge, Lensfield Rd, Cambridge CB2 1EW, UK

SUPPORTING INFORMATION

The supporting information contains the detailed analytic calculations along with schematic diagrams representing the analytical model. It also contains four movies showing reconfigurable quasi-2D colloidal molecules, reconfigurable 3D colloidal molecules, different degrees of flexibility for $N_{max} = 4$ colloidal molecules with size ratios $\alpha = 0.5$ and $\alpha = 0.67$, and molecular dynamics simulation showing the growth of a colloidal molecule with $\alpha = 1$.

ACKNOWLEDGMENTS

This work was financially supported by the European Research Council (ERC) through the starting grant RECONFMATTER (Grant Agreement No. 758383, DJK), the Netherlands Organization for Scientific Research (NWO) through a VIDI grant (LG) and as part of Frontiers of Nanoscience program (NWO/OCW).

# Supporting Information

# Self-assembly dynamics of reconfigurable colloidal molecules

Indrani Chakraborty[†], Daniel J. G. Pearce[‡], Ruben W. Verweij[†], Sabine C. Matysik[†], Luca Giomi[‡], and Daniela J. Kraft*[,†]

[†]   Soft Matter Physics, Huygens-Kamerlingh Onnes Laboratory, Leiden Institute of Physics, PO Box 9504, 2300 RA Leiden, The Netherlands.
[‡]   Institute-Lorentz, Universiteit Leiden, PO Box 9506, 2300 RA Leiden, The Netherlands
*   Corresponding author: kraft@physics.leidenuniv.nl

## Videos

Video S1 on reconfigurable quasi-2D colloidal molecules

Video S2 on reconfigurable 3D colloidal molecules

Video S3 on the different degrees of flexibility for $N_{max}$ = 4 colloidal molecules with size ratios $\alpha$ = 0.5 (left) and $\alpha$ = 0.67 (right).

Video S4 on the molecular dynamics simulation showing the growth of a colloidal molecule with $\alpha$ = 1 and saturation at a valence $N$ = 5.

# Analytic Calculations

## 1 Average valence as function of time

We describe the adding of a colloidal particle to the central particle as a Poisson process. This implies that the average rate at which binding events occur is constant over time, though their actual addition is stochastic. The rate at which a cluster gains an $N^{th}$ particle is constant and proportional to the rate at which the cluster interacts with a potential particle multiplied by the probability that the particle sticks. Since the number of free particles is significantly larger than the number of bound particles, we assume the rate at which the cluster meets other particles is constant and equal to 1/(time for particle to diffuse an average separation). However, the probability that an incoming particle is able to form a bond with the central particle depends on the current number of bound particles. This means the rate at which particles are added to the central cluster changes with the addition of each particle, we dub this process a sequential Poisson process.

Let $t_0$ be the average time between interactions between the cluster and a free particle and $p_N$ be the probability that the incoming particle is able to form a bond with the cluster. Here $p_N$ is a function of the current valence of the cluster, with $p_N$ giving the probability of forming the $N^{th}$ bond. Therefore, in a time interval of length $\Delta t$ the expected number of particles to have interacted and bonded with a cluster of $N - 1$ bound particles is $\lambda = \Delta t p_N / t_0$. Of course, in reality, once a particle has bonded, the value of $p_N$ changes, thus the Poisson process starts again for the addition of the subsequent particle.

In order to derive an analytic description of the sequential Poisson process we start from the cumulative density function (CDF) of a standard Poisson process. This gives us the probability that there have been less than or equal to $k$ events in a time interval in which we would expect $\lambda$ events.

$$P(x \leq k) = e^{-\lambda} \sum_{i=0}^{\lfloor k \rfloor} \frac{\lambda^i}{i!} \tag{1}$$

Therefore, the probability of no events occurring is $P(x \leq 0) = e^{-\lambda}$ and the probability that at least one event has happened is given by $P(x > 0) = 1 - e^{-\lambda}$. Therefore by substituting in for $\lambda$ we define

$$C_N(\Delta t) = 1 - e^{-\Delta t p_N / t_0}. \tag{2}$$

This function is the CDF that a particle successfully binds to a cluster already containing $N-1$ bound particles in a given time interval, $\Delta t$.

The generation of a single large cluster is the result of successive Poisson processes adding a single particle. Let us define the function $G_N(t)$ as the CDF of the $N^{\text{th}}$ particle binding to a cluster which had zero bound particles at time $t=0$. Therefore, the probability density function (PDF) of the same process is given by

$$F_N(t) = \partial_t G_N(t). \tag{3}$$

In order to calculate $G_N(t)$ we must consider the probability of adding an $N^{\text{th}}$ particle given any possible time at which the $N-1^{\text{th}}$ particle was added. This is generally given by

$$G_N(t) = \int_0^t C_N(t-t') F_{N-1}(t') dt'. \tag{4}$$

Since we know that the cluster has zero bound particles at time $t=0$ we can trivially write

$$G_1(t) = 1 - e^{-tp_1/t_0}. \tag{5}$$

This can be used to inductively generate all subsequent functions $G_N$ which are given by

$$G_2(t) = 1 + \frac{p_2 e^{-tp_1/t_0}}{(p_1 - p_2)} + \frac{p_1 e^{-tp_2/t_0}}{(p_2 - p_1)} \tag{6}$$

$$G_3(t) = 1 - \frac{p_2 p_3 e^{-tp_1/t_0}}{(p_1 - p_2)(p_1 - p_3)} - \frac{p_1 p_3 e^{-tp_2/t_0}}{(p_2 - p_1)(p_2 - p_3)} - \frac{p_1 p_2 e^{-tp_3/t_0}}{(p_3 - p_3)(p_3 - p_2)} \tag{7}$$

$$G_4(t) = 1 + \frac{p_2 p_3 p_4 e^{-tp_1/t_0}}{(p_1 - p_2)(p_1 - p_3)(p_1 - p_4)} + \frac{p_1 p_3 p_4 e^{-tp_2/t_0}}{(p_2 - p_1)(p_2 - p_3)(p_2 - p_4)} \qquad (8)$$
$$+ \frac{p_1 p_2 p_4 e^{-tp_3/t_0}}{(p_3 - p_1)(p_3 - p_2)(p_3 - p_4)} + \frac{p_1 p_2 p_3 e^{-tp_4/t_0}}{(p_4 - p_1)(p_4 - p_2)(p_4 - p_3)}$$

$$G_5(t) = 1 - \frac{p_2 p_3 p_4 p_5 e^{-tp_1/t_0}}{(p_1 - p_2)(p_1 - p_3)(p_1 - p_4)(p_1 - p_5)} - \frac{p_1 p_3 p_4 p_5 e^{-tp_2/t_0}}{(p_2 - p_1)(p_2 - p_3)(p_2 - p_4)(p_2 - p_5)} \qquad (9)$$
$$- \frac{p_1 p_2 p_4 p_5 e^{-tp_3/t_0}}{(p_3 - p_1)(p_3 - p_2)(p_3 - p_4)(p_3 - p_5)} - \frac{p_1 p_2 p_3 p_5 e^{-tp_4/t_0}}{(p_4 - p_1)(p_4 - p_2)(p_4 - p_3)(p_4 - p_5)}$$
$$- \frac{p_1 p_2 p_3 p_4 e^{-tp_5/t_0}}{(p_5 - p_1)(p_5 - p_2)(p_5 - p_3)(p_5 - p_4)}.$$

This can be generalized for larger $N$ as

$$G_N(t) = 1 + (-1)^N \sum_{i=1}^{N} Q_i^N e^{-tp_i/t_0} \qquad (10)$$

where the coefficients are given by

$$Q_i^N = \begin{cases} 1, & \text{for } N = 1 \\ \dfrac{\prod_{j \neq i}^{n} p_j}{\prod_{j \neq i}^{n} (p_i - p_j)}, & \text{for } N \neq 1. \end{cases} \qquad (11)$$

The expected valence as a function of time can then be written as the sum of the CDFs, since each Poisson process increases the valence by 1, and hence

$$\langle V(t) \rangle = \sum_{i=1}^{N_{\max}} G_N(t) \qquad (12)$$

where $N_{\max}$ is the maximum valence allowed given the relative size of the inner and outer particles. By definition $G_N(t) = 0$ for $N > N_{\max}$. The prediction of Eq.12 is plotted alongside a simulated sequential Poisson process in Figure S1.

## 1.1 Probability density function of residence time at each valence

The PDF of the time spent at each valence can be written as

$$R_N(\Delta t) = \frac{p_n}{(p_N - p_{N+1})}\left[e^{-\Delta t p_{N+1}/t_0} - e^{-\Delta t p_N/t_0}\right]. \tag{13}$$

This can be derived by recognizing that $R_1 = G_1 - G_2$, which can be generalized for any $N$ by substituting in for $p_N$.

## 1.2 Probability of cluster having a certain valence

This is simply the difference between two subsequent CDFs.

$$P(N,t) = G_N(t) - G_{N+1}(t). \tag{14}$$

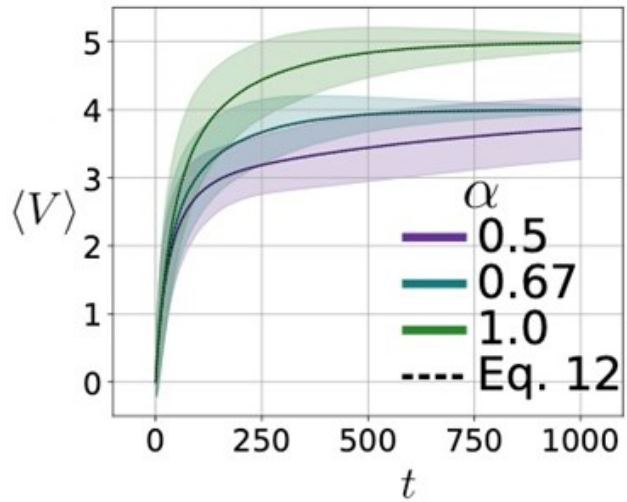

**Figure S1.** Equation 12 plotted along a simulated sequential Poisson process using the values of $p_N$ given in Table I below.

## 2. Probability of $N^{th}$ particle sticking

The probability of each particle sticking is given in the tables below. These are calculated using a statistical approach considering the average available angle for an incoming particle to bind. The first table (Table I) gives the results in 2D in which sedimentation is not considered whereas the second table gives the results when particle sedimentation is considered.

**Table I.** Probability of $N^{th}$ particle sticking for $\alpha = 1.0$, $\alpha = 0.67$ and $\alpha = 0.5$ without considering sedimentation

| N | $\alpha = 1.0$ | $\alpha = 0.67$ | $\alpha = 0.5$ |
|---|---|---|---|
| 1 | 1 | 1 | 1 |
| 2 | 0.670 | 0.594 | 0.543 |
| 3 | 0.381 | 0.258 | 0.182 |
| 4 | 0.179 | 0.062 | 0.014 |
| 5 | 0.047 | - | - |

**Table II.** Probability of $N^{th}$ particle sticking for $\alpha = 1.0$, $\alpha = 0.67$ and $\alpha = 0.5$ considering sedimentation

| N | $\alpha = 1.0$ | $\alpha = 0.67$ | $\alpha = 0.5$ |
|---|---|---|---|
| 1 | 1 | 1 | 1 |
| 2 | 0.670 | 0.584 | 0.509 |
| 3 | 0.381 | 0.242 | 0.137 |
| 4 | 0.179 | 0.050 | 0.001 |
| 5 | 0.047 | - | - |

## 2.1 Calculating the binding probabilities

We take a simple probabilistic approach to recreate the valence curves generated by simulations and experiments. We assume that an incoming particle arrives from a random angle and binds with some probability related to the available space given the number of particles already bound.

The problem is set up as follows, the central particle has radius $\alpha R_0$ and is at fixed position (0,0). The bound particles have radius $R_0$ and their angular positions are given by $\theta$. The bound particles are numbered and have angular positions $\theta_i$ and without loss of generality we set $\theta_1 = -\varphi$ where $\varphi$ is the minimal angle between two bound particles, e.g. $\varphi = \pi/3$ for hexagonal close packed spheres; this sets the origin, see Figure S2a-b.

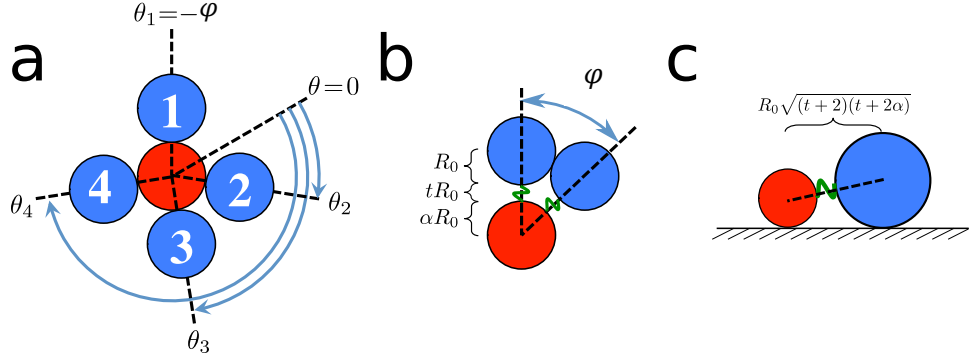

**Figure S2.** (a) Diagram of the variables used in the calculation. (b) Diagram of the definition of $\varphi$. (c) Diagram detailing how the definition of $\varphi$ changes due to the 3D sedimentation of the particles.

## 2.2 Defining the minimum angle $\varphi$ between two bound particles

We define the angle $\varphi$ as the minimum angle between two bound particles. It is essential here to include the tether length in the calculation of $\varphi$, which is defined as $2R_t = t \times R_0$. This is defined by simply applying the cosine rule to the arrangement given in Figure S2b. The angle given here is probably a slight overestimate of the true value as the tether is in part floppy and can have length zero, but we will ignore that here.

$$\cos(\varphi) = 1 - 2(\alpha + t + 1)^{-2} \tag{15}$$

While this estimate of $\varphi$ is good for the simulations, which take place in a strictly 2D environment, for the experiments it must be updated to account for the change in height of the center of the spheres as they sediment onto the surface, see Figure S2c. This is given by

$$\cos(\varphi) = 1 - 2\big((t+2)(t+2\alpha)\big)^{-1}. \tag{16}$$

When $\alpha = 1$ these two equations converge as required.

## 2.3 Finding the binding probabilities

The probability of $p_1$ binding is 100% since all angles are unobstructed in all configurations of the system. Therefore $P(p_1) = 1$. This of course gets more complicated as more particles are added to the system.

### 2.3.1 Adding particle 2

For particle 2, more care is needed. Since particle 1 is already bound, it will be in the way of some trajectories to the central particle. A single particle occludes an arc of $2\varphi$ as the minimum angular separation of two bound particle is $\varphi$, hence the available angular region is $2\pi - 2\varphi$. If we normalise this by the total angle that particle 2 could approach from, we get $P(p_2) = \frac{2\pi - 2\varphi}{2\pi}$. In reality we should integrate over all possible configurations of $p_1$, so we actually get:

$$P(p_2) = \frac{1}{2\pi} \int_0^{2\pi} \frac{2\pi - 2\varphi}{2\pi} d\theta_1 = \frac{2\pi - 2\varphi}{2\pi}. \tag{17}$$

### 2.3.2 Adding particle 3

For the third particle the available space depends on the configuration of the previous two bound particles. We will choose a regime here in which we scan through all configurations by moving the particles one by one from their minimal angular position to their maximal position allowed by the other bound particles. Without any loss of generality, we assume that the initial positions of the particles are given by $\theta'_i = (i - 2) \times \varphi$. We scan through all possible configurations of the system by scanning through the maximal allowed range of $\theta$ for each particle aside from $\theta_1 = -\varphi$ which defines our origin. Scanning through $\theta_1$ which would correspond to rotating our whole system therefore has no effect on available space. This is analogous to scanning through all possible configurations of an abacus. This is an arbitrary choice but makes the problem easier to visualize and the mathematics simpler to formulate.

We now split the problem into two cases which correspond to the potential gaps between the particles. These are outlined below in the schematic in Figure S3.

**Case 1**: $\theta_2 \in [0, 2\pi - 3\varphi]$ (Fig. S3a):

- Probability of $p_2$ being in this region = $(2\pi - 3\varphi)/(2\pi - 2\varphi)$
- Consider the space between particle 2 and 1 in the region $\theta \in [\theta_2 + \varphi, 2\pi - 2\varphi]$.
- Available space for an incoming particle = $2\pi - 3\varphi - \theta_2$

Therefore, this gives a probability of $\frac{2\pi - 3\varphi}{2\pi - 2\varphi} \frac{1}{\int_0^{2\pi - 3\varphi} d\theta_2} \int_0^{2\pi - 3\varphi} \frac{1}{2\pi}(2\pi - 3\varphi - \theta_2)d\theta_2$

**Case 2**: $\theta_2 \in [\varphi, 2\pi - 2\varphi]$ (Fig. S3b):

- Probability of $p_2$ being in this region = $(2\pi - 3\varphi)/(2\pi - 2\varphi)$

- This is in fact the mirror of Case 1, with space in regions $\theta \in [0, \theta_2 - \varphi]$.
- Available space for an incoming particle = $\theta_2 - \varphi$

Therefore, this gives a probability of $\frac{2\pi-3\varphi}{2\pi-2\varphi} \frac{1}{\int_\varphi^{2\pi-2\varphi} d\theta_2} \int_\varphi^{2\pi-2\varphi} \frac{1}{2\pi}(\theta_2 - \varphi)d\theta_2$

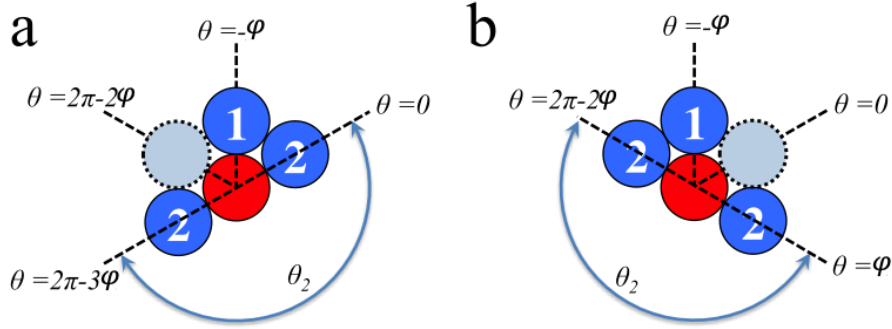

**Figure S3.** Diagram showing the range of $\theta_2$ in which case 1 and case 2 applies. Space for a new particle in regions (a) $\theta \in [\theta_2 + \varphi, 2\pi - 2\varphi]$ and (b) $\theta \in [0, \theta_2 - \varphi]$.

The normalization for the integrations will cancel with the available arc for $p_2$ in all cases and we get:

$$P(p_3) = \frac{1}{2\pi(2\pi - 2\varphi)}\left[\int_0^{2\pi-3\varphi}(2\pi - 3\varphi - \theta_2)\,d\theta_2 + \int_\varphi^{2\pi-2\varphi}(\theta_2 - \varphi)\,d\theta_2\right]. \tag{18}$$

This can be solved to give:

$$P(p_3) = \frac{(3\varphi - 2\pi)^2}{2\pi(2\pi - 2\varphi)}. \tag{19}$$

### 2.3.3 Adding particle 4

Now we must consider the relative positions of three particles, so we have two independent variables and three possible spaces.

**Case 1**: $\theta_3 \in [\varphi, 2\pi - 3\varphi]$:

- Probability of $p_3$ being in this region = $(2\pi - 4\varphi)/(2\pi - 3\varphi)$
- Consider space between particle 1 and 3 in the region $\theta \in [\theta_3 + \varphi, 2\pi - 2\varphi]$.
- Available space for an incoming particle = $2\pi - 3\varphi - \theta_3$.

**Case 2**: $\theta_3 \in [2\varphi, 2\pi - 2\varphi]$ and $\theta_2 \in [0, \theta_3 - 2\varphi]$.

- Probability of $p_3$ being in this region = $(2\pi - 4\varphi)/(2\pi - 3\varphi)$
- Probability of $p_2$ being in this region = $\frac{\theta_3 - 2\varphi}{\theta_3 - \varphi}$.
- Consider space between particle 2 and 3 in the region $\theta \in [\theta_2 + \varphi, \theta_3 - \varphi]$.
- Available space for an incoming particle = $\theta_3 - 2\varphi - \theta_2$.

**Case 3**: $\theta_3 \in [2\varphi, 2\pi - 2\varphi]$ and $\theta_2 \in [\varphi, \theta_3 - \varphi]$:

- Probability of $p_3$ being in this region = $(2\pi - 4\varphi)/(2\pi - 3\varphi)$
- Probability of $p_2$ being in this region = $\frac{\theta_3 - 2\varphi}{\theta_3 - \varphi}$.
- Consider space between particle 1 and 2 in the region $\theta \in [0, \theta_2 - \varphi]$.
- Available space for an incoming particle = $\theta_2 - \varphi$.

The probabilities can be constructed similarly to above to obtain:

$$P(p_4) = \frac{1}{2\pi(2\pi - 3\varphi)} \left[ \int_\varphi^{2\pi - 3\varphi} d\theta_3 (2\pi - 3\varphi - \theta_3) \right. \tag{20}$$

$$+ \int_{2\varphi}^{2\pi - 2\varphi} d\theta_3 \left( \frac{1}{\theta_3 - \varphi} \int_0^{\theta_3 - 2\varphi} d\theta_2 (\theta_3 - 2\varphi - \theta_2) \right)$$

$$\left. + \int_{2\varphi}^{2\pi - 2\varphi} d\theta_3 \left( \frac{1}{\theta_3 - \varphi} \int_\varphi^{\theta_3 - \varphi} d\theta_2 (\theta_2 - \varphi) \right) \right]$$

which can be solved to give:

$$P(p_4) = \frac{4\pi^2 - 18\pi\varphi + 20\varphi^2 + \varphi^2 \log\left[\frac{2\pi - 3\varphi}{\varphi}\right]}{2\pi(2\pi - 3\varphi)}. \tag{21}$$

### 2.3.4 Adding particle 5

This process can be extended to arbitrary many particles added to the system, although with each new particle another layer of integration is added and the problem becomes more complex. We will stop at $p_5$ since we are looking primarily at size ratios $\alpha \leq 1$, so the maximum valence we expect to obtain is 6 for hexagonal close packing.

There are now 3 independent variables which gives rise to four cases.

**Case 1**: $\theta_4 \in [2\varphi, 2\pi - 3\varphi]$

- Consider space between particle 1 and 4 in the region $\theta \in [\theta_4 + \varphi, 2\pi - 2\varphi]$.
- Probability of $p_4$ being in this region = $(2\pi - 5\varphi)/(2\pi - 4\varphi)$
- Available space for an incoming particle = $2\pi - 3\varphi - \theta_4$.

**Case 2**: $\theta_4 \in [3\varphi, 2\pi - 2\varphi]$ and $\theta_3 \in [\varphi, \theta_4 - 2\varphi]$

- Probability of $p_4$ being in this region = $(2\pi - 5\varphi)/(2\pi - 4\varphi)$
- Probability of $p_3$ being in this region = $(\theta_4 - 3\varphi)/(\theta_4 - 2\varphi)$ (given $\theta_4$)
- Consider space between particle 3 and 4 in the region $\theta \in [\theta_3 + \varphi, \theta_4 - \varphi]$.
- Available space for an incoming particle = $\theta_4 - 2\varphi - \theta_3$.

**Case 3**: $\theta_4 \in [3\varphi, 2\pi - 2\varphi]$ and $\theta_3 \in [2\varphi, \theta_4 - \varphi]$ and $\theta_2 \in [0, \theta_3 - 2\varphi]$

- Probability of $p_4$ being in this region = $(2\pi - 5\varphi)/(2\pi - 4\varphi)$
- Probability of $p_3$ being in this region = $(\theta_4 - 3\varphi)/(\theta_4 - 2\varphi)$ (given $\theta_4$)
- Probability of $p_2$ being in this region = $(\theta_3 - 2\varphi)/(\theta_3 - \varphi)$ (given $\theta_3$)
- Consider space between particle 2 and 3 in the region $\theta \in [\theta_2 + \varphi, \theta_3 - \varphi]$.
- Available space for an incoming particle = $\theta_3 - 2\varphi - \theta_2$.

**Case 4**: $\theta_4 \in [3\varphi, 2\pi - 2\varphi]$ and $\theta_3 \in [2\varphi, \theta_4 - \varphi]$ and $\theta_2 \in [\varphi, \theta_3 - \varphi]$

- Probability of $p_4$ being in this region = $(2\pi - 5\varphi)/(2\pi - 4\varphi)$
- Probability of $p_3$ being in this region = $(\theta_4 - 3\varphi)/(\theta_4 - 2\varphi)$ (given $\theta_4$)
- Probability of $p_2$ being in this region = $(\theta_3 - 2\varphi)/(\theta_3 - \varphi)$ (given $\theta_3$)
- Consider space between particle 1 and 2 in the region $\theta \in [0, \theta_2 - \varphi]$.
- Available space for an incoming particle = $\theta_2 - \varphi$.

This can be combined to give the formula:

$$P(p_5) = \frac{1}{2\pi(2\pi - 4\varphi)} \Bigg[ \int_{2\varphi}^{2\pi-3\varphi} d\theta_4 (2\pi - 3\varphi - \theta_4) \qquad (22)$$

$$+ \int_{3\varphi}^{2\pi-2\varphi} d\theta_4 \left( \frac{1}{\theta_4 - 2\varphi} \int_{\varphi}^{\theta_4-2\varphi} d\theta_3 (\theta_4 - 2\varphi - \theta_3) d\theta_3 \right) d\theta_4$$

$$+ \int_{3\varphi}^{2\pi-2\varphi} d\theta_4 \left( \frac{1}{\theta_4 - 2\varphi} \int_{2\varphi}^{\theta_4-\varphi} d\theta_3 \left( \frac{1}{\theta_3 - \varphi} \int_{0}^{\theta_3-2\varphi} d\theta_2 (\theta_3 - 2\varphi - \theta_2) \right) \right)$$

$$+ \int_{3\varphi}^{2\pi-2\varphi} d\theta_4 \left( \frac{1}{\theta_4 - 2\varphi} \int_{2\varphi}^{\theta_4-\varphi} d\theta_3 \left( \frac{1}{\theta_3 - \varphi} \int_{\varphi}^{\theta_3-\varphi} d\theta_2 (\theta_2 - \varphi) \right) \right) \Bigg]$$

which can be solved to give:

$$P(p_5) = \frac{4\pi^2 - 24\varphi\pi + 35\varphi^2 + 2\varphi^2 \log\left[\frac{2\pi - 4\varphi}{\varphi}\right] + \frac{\varphi^2}{4}\left(\log\left[\frac{2\pi}{\varphi} - 4\right]\right)^2}{2\pi(2\pi - 4\varphi)}. \qquad (23)$$

We can now substitute in $\varphi$ for the different cases we are looking at and find the probabilities given in the table earlier.

## 2.4 Generating function for arbitrary valence

It is possible to write a generating function for the probability of addition of particles to a system with arbitrary currently bound particles. First, we need to introduce some new notation. We define $\theta_i'$ as the initial position of particle $i$, which under our current set up is $\theta_i' = (i - 2)\varphi$. For a system with $N$ currently bound particles, we only use $N - 1$ free parameters as $\theta_1 = -\varphi$. In order to accommodate the periodicity in the system we must define $\theta_{N+1} = 2\pi - \varphi$ and the initial positions of these as $\theta_1' = -2\varphi$ and $\theta_{N+1}' = (N - 1)\varphi$.

This equation will obviously only hold when the minimum angle between adjacent particles is small enough to accommodate all the particles currently bound, i.e. $(N + 1)\varphi < 2\pi$, otherwise it is impossible to add an additional particle.

The generating function for the probability of adding particle $N + 1$ is then:

$$P(p_{N+1}) = \frac{1}{2\pi} \sum_{i=1}^{N} \left[ \overrightarrow{\prod_{j=i+1}^{N+1}} g_j \circ \right] \frac{1}{\theta_{i+1} - \theta_{i+1}'} \int_{\theta_{i'}}^{\theta_{i+1}-2\varphi} (\theta_{i+1} - 2\varphi - \theta_i) \, d\theta_i \quad (24)$$

where $g_j$ is a function defined as follows:

$$g_j(f) = \begin{cases} f, & \text{if } j = N+1 \\ \frac{1}{\theta_{j+1} - \theta_{j+1}'} \int_{\theta_{j'}+\varphi}^{\theta_{j+1}-\varphi} f \, d\theta_j, & \text{otherwise} \end{cases} \quad (25)$$

and $f_i$ is a function defined as

$$f_i = \frac{1}{\theta_{i+1} - \theta_{i+1}'} \int_{\theta_{i'}}^{\theta_{i+1}-2\varphi} (\theta_{i+1} - 2\varphi - \theta_i) \, d\theta_i \quad (26)$$

where the order of the product in Eq. 24 is indicated by the arrow, hence the product is a function which acts on the final integral

$$\begin{aligned} & \left[ \overrightarrow{\prod_{j=i+1}^{N+1}} g_j \circ \right] \frac{1}{\theta_3 - \theta_3'} \int_{\theta_2'}^{\theta_3-2\varphi} (\theta_3 - 2\varphi - \theta_2) \, d\theta_2 \\ &= g_5 \left( g_4 \left( g_3 \left( \frac{1}{\theta_3 - \theta_3'} \int_{\theta_2'}^{\theta_3-2\varphi} (\theta_3 - 2\varphi - \theta_2) \, d\theta_2 \right) \right) \right) \\ &= \frac{1}{\theta_6 - \theta_6'} \int_{\theta_5'+\varphi}^{\theta_6-\varphi} \frac{1}{\theta_5 - \theta_5'} \int_{\theta_4'+\varphi}^{\theta_5-\varphi} \frac{1}{\theta_4 - \theta_4'} \int_{\theta_3'+\varphi}^{\theta_4-\varphi} \frac{1}{\theta_3 - \theta_3'} \int_{\theta_2'}^{\theta_3-2\varphi} (\theta_3 - 2\varphi - \theta_2) \, d\theta_2 d\theta_3 d\theta_4 d\theta_5. \end{aligned} \quad (27)$$